\documentclass[manuscript]{aastex}
\usepackage{amsmath}

\slugcomment{}

\shorttitle{IRAC constraints on Ho~IX }
\shortauthors{Allue et al.}

\begin{document}

\title{{\it Spitzer} IRAC observations of IR excess in Holmberg~IX X-1:  A circumbinary disk or a variable jet? }

\author{R. P. Dudik\altaffilmark{1}, C. T. Berghea\altaffilmark{1}, T. P. Roberts\altaffilmark{3}, F. Gris{\'e}\altaffilmark{4}, A. Singh\altaffilmark{2}, R. Pagano\altaffilmark{5}, L. M. Winter\altaffilmark{6}}

\altaffiltext{1}{U.S. Naval Observatory (USNO), 3450 Massachusetts Avenue NW, Washington, DC 20392, USA}
\altaffiltext{2}{George Mason University, 4400 University Drive, Fairfax, VA 22030}
\altaffiltext{3}{Centre for Extragalactic Astronomy, University of Durham, South Road, Durham DH1 3LE, UK}
\altaffiltext{4}{Observatoire astronomique de Strasbourg, Universit{\'e} de Strasbourg, CNRS, UMR 7550, 11 rue de l'Universit{\'e}, F-67000 Strasbourg, France}
\altaffiltext{5}{St. John's College, 60 College Ave, Annapolis, MD 21401, USA}
\altaffiltext{6}{Atmospheric and Environmental Research, 131 Hartwell Avenue, Lexington, MA 02421, USA}

%%\affil{United States Naval Observatory, Washington, DC 20392}

\email{rachel.dudik@usno.navy.mil}

\begin{abstract}

We present {\it Spitzer} Infrared Array Camera (IRAC) photometric observations of the Ultra-luminous X-ray Source (ULX, X-1) in Holmberg IX.  We construct a spectral energy distribution (SED) for Holmberg IX X-1 based on published optical, UV and X-ray data combined with the IR data from this analysis.  We modeled the X-ray and optical data with disk and stellar models, however we find a clear IR excess in the ULX SED that cannot be explained by fits or extrapolations of any of these models. Instead, further analysis suggests that the IR excess results either from dust emission, possibly from a circumbinary disk or from a variable jet.  
\end{abstract}

\keywords{black hole physics --- galaxies: individual (Holmberg IX) --- X-rays: binaries}

\section{INTRODUCTION}

Ultra-luminous X-ray Sources (ULXs) are unusually bright X-ray sources, with L$_X$ $>$ $10^{39}$ erg s$^{-1}$, which is approximately the Eddington limit for a 10~M$_{\odot}$ object \citep[e.g.][]{fen11}. Even though ULXs were discovered roughly three decades ago, the underlying mechanism that produces the powerful X-ray emission in many of these objects remains uncertain. However, after years of multi-wavelength observations, only a few ULXs remain strong candidates for the Intermediate Mass Black Hole (IMBH) scenario, such as HLX-1 in the galaxy ESO 243-49 \citep{farr09, webb12} and M82~X-1 \citep{str03}, while many others show indications of super-Eddington accretion and/or beaming from a stellar mass black hole \citep[sMBH, e.g.][]{sto06, king08, pou07, sor08, ber08, gla09, sut13, liu13, mot14, weng14, mid15} or a neutron star \citep{bach14}. Some galactic binaries are known to show super-Eddington luminosities, such as GRS~1915+105 \citep{fen04} and V4641~Sgr \citep{rev02}, but not persistently as ULXs do. However with the exception of a few distinct cases \citep[see][]{mid13, liu13, mot14} conclusive evidence on the true nature of many ULXs remains unclear.

The ULX in Holmberg IX (Ho~IX X-1 from now on) is one such enigmatic source. The ULX, also known as Holmberg~IX~X-1 or M81~X-9, has been well studied in the X-rays and optical since its discovery by the Einstein Observatory \citep{fabb88}. In this paper we assume that the ULX is located at a distance of 3.6~Mpc \citep{ger11}. Perhaps one of the most notable features of this ULX is the massive nebula in which it resides.  Stretching ~300 pc $\times$ 400 pc in size, the nebula was first discovered and associated with the ULX by \citet{mill94} and \citet{lap01} respectively.  The ultimate power source for the bubble nebula has been a matter of some debate.  Some explanations for its origin include 1) the combined strength of multiple O-stars and supernovae from the OB association in close proximity to the ULX, 2) a large hypernova event, or 3) winds and/or jets emanating from the ULX \citep[e.g.][]{pak08, abo08, abo07,ram06,pak02,pak03}.

The suggestion that winds or jets might power the massive nebular bubble around Ho~IX X-1 is in line with the theory that many ULXs are similar to the famous supercritically accreting X-ray binary in our own Galaxy, SS~433, rather than an IMBH. Indeed in the case of Ho~IX X-1, the other two scenarios have ostensibly been ruled out as plausible explanations for the nebula distribution.  For instance, \citet{ram06} showed that the energy released by six supernovae in the vicinity of the ULX is not sufficient to reproduce the energy of the expanding nebular shell, ruling out (1) above.  \citet{pak08} similarly find that the mechanical energy required to produce the nebula must result from a cluster that is more than two orders of magnitude larger than the observed stellar association.  They also suggest that the period of time needed for mass transfer to begin in Ho~IX X-1 is far shorter than required for a hypernova event to explain the birth of the ULX, ruling out (2). \citet{pak08} suggest that winds and/or jets instead are more likely to power the bubble. \citet[Figure~10]{ber10a} support this finding, when they use shock models to describe the high excitation optical lines seen in Ho~IX X-1.  Finally, \citet{abo07} find radial velocity gradients that support the winds and/or jets explanation.

Many authors have noted the striking resemblance of Ho~IX X-1 to SS 433.  For instance, \citet{fab15} note that the optical spectrum of Ho~IX X-1 is similar to wind-dominated objects like SS 433.  In an X-ray study, \citet{lua16} see spectral variability patterns that are consistent with precession of the angle to the line-of-sight of the rotation axis of the ULX.  This is very similar to what is seen in SS 433 as well.   These scenarios hint that the processes powering emission in Ho~IX X-1 and SS 433 may be similar, so much so that the two appear to be of the same family.

If the Ho~IX X-1 nebula is indeed powered by jets as is the case with its SS 433 `cousin', direct evidence of these jets has yet to surface.  However infrared photometry - an underused tool in the field of ULX astronomy - has the potential to produce a wealth of continuum information about the ULX that can be used to directly characterize the ULX SED.  This has been done only once before for ULXs: in a study of Holmberg~II X-1. In this study \citet{ber10a, ber10b} used Spitzer mid-infrared data to constrain the ULX SED and found the data to be consistent with a broken power law typical of jet emission. The model was substantiated when \citet{cseh14} used the Karl G. Jansky Very Large Array (JVLA) to image the ULX and found signatures of jets emanating from the ULX.  On the other hand, infrared photometry has been widely used in the X-ray binary community to uncover excess from irradiated disks, circumbinary disks, and dusty shells from companions stars \citep{rah10, mun06}.  Thus Spitzer infrared photometry can be a powerful tool for uncovering  structures and environments of ULXs, that optical and X-ray data cannot. 

In this paper we present Spitzer IRAC images of the ULX in Holmberg IX.  Combining these images with multi-wavelength optical, UV, and X-ray data from the literature, we model the Ho~IX X-1 SED to determine the source of the infrared emission in this ULX. In Section 2 we provide the details of the Spitzer IRAC data analysis.  In Section 3 we provide the fits to the full SED. In Section 4 we discuss the origin of the infrared emission. Finally we summarize our results in Section 5.  

\section{IRAC OBSERVATIONS AND DATA ANALYSIS}

For this analysis, we used two sets of archival Spitzer Infrared Array Camera (IRAC) data of the dwarf galaxy Holmberg IX taken on November 15, 2007 (Astronomical Observation Request, AORs 22537472 and 22537728). IRAC is a four-channel camera that provides 5.12 $\times$ 5.12 arcmin images at 3.6, 4.5, 5.8 and 8.0 microns.  We processed the data both manually and using the Post-processed Basic Calibration Data (PBCD) available for download from the archive, but found no difference between the two methods.  

We used the Spitzer MOsaicker and Point Source Extractor (MOPEX) APEX package\footnote{http://irsa.ipac.caltech.edu/data/SPITZER/docs/dataanalysistools/tools/mopex/mopexusersguide} to extract the photometric fluxes from the four channels. We followed the standard extraction procedure and conversion factors provided in the IRAC Instrument Handbook\footnote{http://irsa.ipac.caltech.edu/data/SPITZER/docs/irac/iracinstrumenthandbook/IRAC\_Instrument\_Handbook.pdf}. The pipeline mosaics for IRAC are accurate to within 20\%\footnote{http://irsa.ipac.caltech.edu/data/SPITZER/docs/irac/iracinstrumenthandbook/74/} of the flux.   All data calibration information, including background subtraction, are available in the footnoted links as well as in \citet{mak05}, which contains the exact point source extraction procedure from MOPEX.  To ensure our IR data were extracted from the same region as the Optical and X-ray data, we aligned the IRAC images with the {\it Hubble Space Telescope} (HST) Advance Camera for Surveys (ACS) F555W/V image using 11 common sources in both (HST ACS observation ID GO-9796 in Table 1). We find that the sources in the two images align to within 0.11 arcsec RMS. 

Table~1 provides fluxes extracted for the two sets of data and Figure~1 shows false-color images of the ULX in 3.6, 4.5, and 8.0 microns compared with a standard HST ACS image of the ULX. We also used a SUBARU H$_\alpha$ image to overplot the contours of the bubble nebula.  As the fluxes in Table~1 show, the observations at 5.8 and 8.0 microns show a weak detection in one case (AOR 22537472) and an upper limit in another (AOR22537728) for a detection threshold of 4-sigma.  Indeed both observations are right on the line between detection and a non-detection for the 4-sigma threshold.  Because the two sets of observations are identical in exposure time and because they are contiguous (i.e. one observation was taken right after the other), we chose to combine the two sets of observations to reduce the noise error on the measurements.  The resultant fluxes are also provided in Table~1.  The enhanced signal to noise in the combined measurement suggests a statistically significant detection at 5.8 and 8.0 microns that can be used to characterize the IR emission.  

\section{SPECTRAL ENERGY DISTRIBUTION OF HOLMBERG IX X-1}

To generate a full Spectral Energy Distribution (SED) of  Ho~IX X-1, we used optical, UV, and X-ray data from the literature \citep{gris11,ber12,lua16}. A summary of all archival data used for this analysis is presented in Table~\ref{table1}.  The optical data comes primarily from HST and the reduction strategy for that data can be found in \citet{gris11}.  The UV data primarily comes from Optical Monitor on board {\it XMM-Newton} and the data analysis strategy for this data can be found in \citet{ber12}.  Finally the X-ray data comes from {\it XMM-Newton} and the reduction strategy for these data can be found in \citet{lua16}.  We used the extinction E(B-V) = 0.26 adopted by \citet{gris11} to correct for absorption in the UV, optical and IR.  The extinction curves used for this correction are taken from \citet{card89}. 

This multi-wavelength data sample spans 5 years.  We note that Ho~IX  X-1 is known to show X-ray variations of factors of 4-5 over days to years \citep{vier10, lua16}.   Unfortunately, the only available multi-wavelength dataset that was taken of this object includes X-ray data from {\it Chandra} that suffers from severe pile-up and is therefore unusable for the X-ray fitting portion of this analysis \citep{gris11}.  However we were able to use this {\it Chandra} data set to establish an estimate for the approximate X-ray flux during the same epoch as the HST data used here.   Indeed we find that the Chandra flux is very close to the mean between the two sets of {\it XMM} X-ray observations.  Therefore, in the following, note that the actual HST-epoch X-ray flux is likely somewhere between the two {\it XMM} X-ray data sets and that the plotted data set provided here is {\it not} simultaneous.   As the following sections will demonstrate, the multi-epoch nature of these data makes generating a good fit difficult.

X-ray spectra exist in the archives from {\it XMM-Newton}, {\it Swift}, and {\it NuSTAR}. \citet{lua16} found two main types of spectra: one hard ultra-luminous (HUL)-like and one more disk-like \citep[see also][]{sut13}. We selected the {\it XMM-Newton} datasets 0112521101 and 0693851701 (refered to as obs 1 and obs 2 respectively from here forward), which represent each of the two, typical ULX spectra regimes seen in this object.

\subsection{Spectral Fitting}

We used XSPEC version 12.7.1 to model the two X-ray spectra of Ho~IX X-1 between 0.3 and 10 keV.  The X-ray data were provided by and processed per the prescription of \citet{lua16}.  The X-ray data were fit with three models:  multi-color disk (MCD) + comptonization model (compTT), Self-Irradiated Funnel model (SIRF), and an irradiated disk model (DISKIR).   Note that the former two were fit to the X-ray data only, while the latter (DISKIR) was fit to both the X-ray and Optical data since that made the most physical sense for this model as described in the following.

We start with a model typically used to fit ULXs, with an accretion disk component (MCD in XSPEC) and the CompTT model to account for the Comptonization in a disk-corona geometry \citep[e.g.][]{gla09, wal14}. We obtain good fits for both spectra. The parameter fit results are shown in Table~\ref{table2}.

The Self-IrRadiated Funnel model (SIRF in XSPEC), describes the emission from a ``supercritical funnel'', similar to SS~433 \citep{abo09}.  The SIRF model provides acceptable fits for both sets of X-ray spectra. The accretion rate was fixed at 350 $\times$ Eddington, the funnel opening angle was set to 30 degrees and the inclination angle was set to zero. The inner radius for this model is calculated in units of the ``spherisation radius'' \citep{abo09}. In order to better compare them with the other models we convert them in units of km using the luminosity. We obtained estimates for the bolometric luminosities from the fitted X-ray models plus the jet model fit to the IRAC data (see Section 4). Then we used the accretion rate to estimate black hole masses of 41~M$_{\odot}$ and 107~M$_{\odot}$ for the first and the second observation, respectively. The spherization radius can then be calculated \citep{abo09} for each black hole mass. Finally we convert the fitted radii in the SIRF model to km. These are listed in column 5 in Table~\ref{table2}.

The DISKIR model is particularly useful when optical and UV data are available in addition to the X-ray spectrum. This is because many ULXs detected in the optical/UV show SEDs well fit by an irradiated disk \citep{ber10a, ber12, gris12}.  In this case it makes sense to fit the X-ray data together with the optical data, since the irradiation portion describes a large fraction of the optical/UV emission.  Indeed an irradiated disk was found to have an important contribution in the optical in other ULXs \citep[e.g.][]{ber12}.  Following Gierlinski et al. (2008, 2009) and Berghea \& Dudik (2012) we set the fraction of the flux thermalized in the inner disk at 0.1 and the radius of the Compton irradiated disk at 1.1 of the inner disk radius.  A good fit requires we set a large Comptonized luminosity ($\geq$ unilluminated disk luminosity), and an outer disk of radius 1000 R$_{in}$ (where R$_{in}$  is the inner disk radius), but this parameter is not well constrained. The fits are not acceptable statistically, but we obtained estimates for the irradiated flux fraction. It is large for both X-ray obs 2 (HUL-like, $>$6.8\%), and X-ray obs 1 (disk-like, $>$5.7\%).  Typical values of 2-4\% have been found for previous fits of ULX X-ray/optical datasets with the DISKIR model \citep{tao12, gris12, ber12}. 

\citet{sut14} used an improved model for irratiated disks (one that accounts for the color-temperature correction) to fit a sample of ULXs with disk-like spectra.  In this study, \citet{sut14} compared the new DISKIR model results with a more traditional DISKIR model fit.  They found that for most objects the reprocessing fraction is ten times lower than predicted by traditional DISKIR models that do not include the color-temperature corrections, and that these new reprocessing fractions are closer to what is observed in Galactic binaries.  However for one of the brighter objects in that sample (NGC1313 X-2) they found a similar reprocessing fraction to ULXs where the spectrum may be more HUL-like.  They interpreted the higher reprocessing fraction in the brighter ULX as originating in a higher scattering fraction onto the outer disk from the extensive wind thought to be driven off the disk; the high reprocessing fraction in Ho IX X-1 is consistent with this interpretation.

\subsection{Extending the models to Longer Wavelengths}

In the following we construct the SED by extending the X-ray models we fit in the previous section to the optical/UV and IR data.  

{\bf Extrapolated SIRF Model:} The extrapolated SIRF models are plotted in Figure~\ref{sed}. Neither of these X-ray models fit the optical data well, however the optical data is located between the two SIRF models. When the X-ray spectrum is disk-like, it over-predicts the optical data by a factor of 6.2.  When the X-ray spectrum is HUL-like it under-predicts the optical data by a factor of 3.2. Taking into account the variability of the Ho~IX X-1 and the multi-epoch nature of this data set we conclude that a funnel model is capable of reproducing the optical and X-ray data.  However Figure~\ref{sed} clearly shows that the SIRF model does not fit any of the IR data.   Indeed if this SIRF model were the appropriate model for the optical and X-ray data, an additional component would be needed to explain the IR excess.  

{\bf Extrapolated Disk Models:}   As expected, simple disk models without irradiation are too faint in the optical. Our MCD+CompTT models shown in Figure~\ref{sed} are more than an order of magnitude lower than the Hubble data if they are extrapolated into the optical. Following \citet{gris11} we added a B0~Ib star to fit the optical photometry (black line in Figure~\ref{sed}) in this case, which resulted in a good fit to the data from X-ray to optical.  The optical data are fit so well by the stellar model, that an IR excess is clearly present at wavelengths longer than the H-band measurement.  Thus an additional component is needed to explain the IRAC data for the extrapolated disk model as well.  

{\bf DISKIR model:}   Per Section 3.1 we began fitting the DISKIR data to the optical + X-ray data since it makes physical sense to do so when optical data is available.   This extrapolated DISKIR model, described in Section 3.1, was a very good fit to the optical data but unfortunately not statistically acceptable overall for either X-ray data set.  Here too, the DISKIR model is also not a good fit the IRAC data.  

Based on these fits and the multi-epoch nature of this dataset, we conclude that all three models are plausible models for the optical/UV/X-ray data in Ho~IX X-1.  To definitively rule out any one model would require good, simultaneous optical and X-ray data (Gris\'e et al. 2016, in prep).   However, most importantly,  none of these models fit the IR excess we see in Ho~IX X-1, suggesting another mechanism is responsible for this emission.

\section{ORIGIN OF THE IR EMISSION IN HOLMBERG IX X-1}

As Figure~\ref{sed} shows, there is a clear IR excess in Ho~IX X-1 that cannot be explained by the optical or X-ray models.  The IR excess could be due to contamination from other stars within the 2 arcsecond extraction region, however based on the B0~Ib stellar model in Figure 2 (which is also the brightest star in the 2 arcsecond extraction circle), we expect the contribution from all stars in that field to be negligible in the IR since their collective SEDs drop sharply at IR wavelengths. 

There are four possible sources for the IR excess:  1) the irradiated disk/the companion star, 2) a heated dust shell, 3) a circumbinary disk such as those seen in X-ray binaries, or 4) a jet. Because none of the models from the previous sections produce sufficient emission in the IR to replicate our IRAC data, the latter three options are the only plausible emission mechanisms. We note that some fraction of the IR emission will also result from contamination from the bubble nebula.   \citet{gris11} found some red excess in the I-band based on the stellar model fits to the HST data.  However the HST images from this study suggest the I-band emission is not coming from nebular contamination, especially in the central location \citep{gris11}.  In addition the H-band fits the stellar continuum very well and does not show a similar excess.  We explore the three remaining options for the excess below.

{\bf Heated Dust Shell:}  Emission from a heated dust shell is one explanation for the IR excess seen in Ho~IX X-1.  \citet{rah10} found evidence for a dust component with temperature $\sim$400~K from  the Galactic X-ray binary GRS 1915+105.  GRS1915+105 is a microquasar  that has a red giant as a donor \citep{gre01}.  \citet{rah10} suggests that the dust component could be related to the dusty shell often found surrounding red giants. We fit a spherical blackbody model to the Ho~IX X-1 IRAC data.  This blackbody, shown in Figure~\ref{sed},  gives a dust temperature of $\sim$ 1100~K and a radius of $\sim$ 1400~R$_{\odot}$.   However, the fit is quite poor, the $\chi^2_{\nu}$ being much greater than two, thereby precluding error estimates for the parameters. The dust temperature is normal for dusty shells around red giants, however if the companion star is a blue supergiant, as the optical HST data suggests, it is unlikely to be surrounded by dust.  In addition this blackbody emission does not fit the 5.8 and 8.0 $\mu$m data well.  We therefore rule out this option as a viable model for the infrared emission in the ULX.   

{\bf Circumbinary Disk:}  An alternative explanation for a dust component is the circumbinary disk proposed by \citet{per10}, \citet{rah10} and \citet{mun06} to explain the IR emission in X-ray binaries (SS~433, GRS1915+105, A0620-00 and XTE J1118+480). Here the circumbinary disk is distinct from the accretion disk described in Section 3, and is composed of material that may have been lost through the L2 point \citep[e.g. SS433, see][]{per10}. If Ho~IX X-1 is more like these X-ray binaries, we might expect to see such a circumbinary disk in the infrared. In this case the black body model that we fit to the infrared data would represent emission from the inner circumbinary disk, while the redder 5.8 and 8.0 $\mu$m excess is likely emanating from the outer portion of the circumbinary disk as lower temperature blackbody emission.  The four IRAC data points that comprise the IR excess in Ho~IX X-1 are not sufficient to fit a complicated circumbinary disk model \citep[e.g.][]{hil15, ake07}, however in its simplest form, the circumbinary disk model is very similar to an irradiated disk model, only the source of illumination in the former case is the accretion disk and the star.  We therefore fit a very simple p-free disk model to the four IRAC points.  As can be seen from Figure 2, the data are consistent with this simple p-free disk (${p = 0.64^{+0.09}_{-0.06}}$), but the disk vastly over-predicts the H-band flux and the temperature for the fit is unconstrained ($T > 1680 K$).  We conclude that if a circumbinary disk is responsible for the IRAC emission in this object then the optical and/or IR emission must be variable.  

{\bf Jet Emission:}  In the jet emission scenario we expect a broken power law with a break in the IR and we attempted to fit the combined IRAC and HST data with such a model.  Extinction E(B-V) was fixed at 0.26 and the break energy at 0.35~eV (approximately at the 3.6 $\mu$m IRAC band, see Figure~\ref{sed}). For comparison the break is at 0.12~eV for Cygnus X-1 and 0.48 for GX~339-4 \citep{rah11}. Inspection of the fit clearly indicates that the HST H-band flux is prohibitively low for a reasonable power law model.  The emission in H-band fits the stellar model so well that any additional contribution from a jet-dominated power law would either severely over predict the observed emission or suggest a power law slope that is unrealistically steep \citep{bla79, fal95}.  Indeed the spectral index of the fit in Figure~\ref{sed} is $-4.3$ in the optical/IR compared with typical indices of $-0.4$ to $-1.0$ \citep{bla79, fal95}.  We conclude from this that either a) the IR emission from Ho~IX X-1 is {\it not} jet-dominated or b) that the radio ejecta are transient as is thought to be the case with Ho~II X-1 \citep{cseh14}.  The multi-epoch nature of the HST and IR data results in a degeneracy in the models that can only be broken by a series of radio monitoring activities designed to detect the jet emission when it is in its most luminous state.

Based on the analysis presented in the previous sections, the IR emission in Ho~IX X-1 suggests that the system contains a circumbinary disk much like those detected in X-ray binaries (e.g SS~433, GRS1915+105, A0620-00 and XTE J1118+480) or emission from a variable jet such as that seen with Ho~II X-1 \citep[e.g.][]{cseh14}.   In the transient jet scenario, simultaneous IR and optical data might uncover a more reasonable power-law slope and H-band emission that is significantly higher than observed in the data set presented here.  We also find that the IR data is not well fit by a single black body, such as that expected from a dusty shell.  However, in this case the simultaneity of the data will not significantly improve the fit, since the single black body fits neither the optical data nor the longer wavelength IR data as well.  The degeneracy between the circumbinary disk and transient jet theory underscores the need for simultaneous observations when observing ULX structure and environments.  In this case, very sensitive radio observations of Ho~IX X-1 may solidify these findings and help detect or constrain the power law break and slope needed to confirm jet activity.  Indeed such radio observations have recently been obtained using the JVLA B-array and will be published in a follow-on study to this paper. However, as this study also indicates, deep radio {\it monitoring} observations with JVLA are also critical to providing information about the IR excess in this source in the event that the jet (if one exists) is variable.    

\section{CONCLUSIONS}

Using {\it Spitzer} IRAC observations of Ho~IX X-1, we have constructed an SED of the ULX. Two contiguous IRAC observations of Ho~IX X-1 were made.  The datasets at 5.8 and 8.0 microns are at the sensitivity limit of the IRAC instrument, however combining these measurements yields a statistically significant detection in both bands.  The combined measurements coupled with detailed fits to previous optical/UV and X-ray data of Ho~IX X-1 suggest that the IR excess in this object is due either to a circumbinary disk such as those seen in SS~433 and other standard X-ray binaries or a variable jet such as that seen in Ho~II X-1 by \citet{cseh14}. Future high-sensitivity radio monitoring observations would be needed to break the degeneracy between the two models and determine if the IR excess seen in Ho~IX X-1 results from either mechanism.

\acknowledgments
We thank the anonymous referee for the very helpful suggestions that have greatly improved this paper. The authors thank Luangtip et al. for providing the X-ray data presented here and for their useful discussions. We also thank Michel Hillen, Hans Van Winckel, and Rachel Akeson for the useful discussions concerning circumbinary disk models and their application to this dataset. TPR acknowledges financial support from STFC as part of the consolidated grant ST/L00075X/1. This work is based on observations obtained from multiple telescope facilities including 1) {\it XMM-Newton}, an ESA science mission with instruments and contribution directly funded by ESA Member States and NASA, 2) The NASA/ESA {\it Hubble Space Telescope} and obtained from the Hubble Legacy Archive, which is a collaboration between the Space Telescope Science Institute (STScI/NASA), the Space Telescope European Coordinating Facility (ST-ECF/ESA) and the Canadian Data Centre (CADC/NRC/CSA), and finally, 3) The NASA {\it Spitzer Space Telescope} which is managed by the Jet Propulsion Laboratory with data obtained from the {\it Spitzer Heritage Archive} which is maintained by the Spitzer Science Center (SSC), located on the campus of the California Institute of Technology and part of NASA's Infrared Processing and Analysis Center (IPAC)  The Spitzer Space Telescope is a NASA mission managed by the Jet Propulsion Laboratory. This website is maintained by the Spitzer Science Center, located on the campus of the California Institute of Technology and part of NASA's Infrared Processing and Analysis Center.

%%%%%%%%%%%%%%%%%%%%%%%%%%%%%%%%%%%%%%%%%%%%%%%%%%%%%%%%%%%%%%%%%%%%%%%%%%%%%%%%%%%%%%%%%%%

\begin{deluxetable}{ccccccc}
\tablecolumns{6}  
\tablewidth{0pt} 
\tabletypesize{\tiny}
\setlength{\tabcolsep}{0.05in}
\tablenum{1} %\\               
\tablecaption{Multiwavelength Data\label{table1}}   
\tablehead{
\colhead{Instrument} & \colhead{Obs ID} & \colhead{Filter} & \colhead{Obs. Date} & \colhead{Flux density}  & \colhead{Abs. }  & \colhead{Corrected} \\
\colhead{} & \colhead{} & \colhead{} & \colhead{} & \colhead{($\mu$Jy)} & \colhead{(A$_{\lambda}$)} & \colhead{($\mu$Jy)} \\
\colhead{(1)} & \colhead{(2)} & \colhead{(3)} & \colhead{(4)} & \colhead{(5)} & \colhead{(6)} & \colhead{(7)}  \\
}
\startdata

Spitzer IRAC  &  22537472  &	3.6 & 	2007 Nov 15 & 	4.99$\pm$0.23 & 0.049 & 5.22$\pm$0.24 \\
Spitzer IRAC  &  22537472  &	4.5 & 	2007 Nov 15 & 	4.94$\pm$0.36 & 0.046 & 5.15$\pm$0.38 \\
Spitzer IRAC  &  22537472  &	5.8 & 	2007 Nov 15 & 	6.98$\pm$1.5 & 0.042 & 	7.25$\pm$1.57 \\
Spitzer IRAC  &  22537472  &	8 & 	2007 Nov 15 & 	7.56$\pm$1.8 &	0.039 & 7.65$\pm$1.87 \\
Spitzer IRAC  &  22537728  &	3.6 & 	2007 Nov 15 & 	4.89$\pm$0.23 &	0.049 & 5.12$\pm$0.24 \\
Spitzer IRAC  &  22537728  &	4.5 & 	2007 Nov 15 & 	5.09$\pm$0.36 & 0.046 & 5.31$\pm$0.37 \\
Spitzer IRAC  &  22537728  &	5.8 & 	2007 Nov 15 & 	$<$4.41 & 0.042 & $<$4.59 \\
Spitzer IRAC  &  22537728  &	8 & 	2007 Nov 15 & 	$<$5.45 & 0.039 & $<$5.65 \\
Combined IRAC  &  $\cdots$  &	3.6 & 	2007 Nov 15 & 	4.94$\pm$0.16 & 0.049 & 5.17$\pm$0.17 \\
Combined IRAC  &  $\cdots$  &	4.5 & 	2007 Nov 15 & 	5.02$\pm$0.25 & 0.046 & 5.23$\pm$0.27 \\
Combined IRAC  &  $\cdots$  &	5.8 & 	2007 Nov 15 & 	5.70$\pm$1.01 & 0.042 & 5.92$\pm$1.10 \\
Combined IRAC  &  $\cdots$  &	8 & 	2007 Nov 15 & 	6.51$\pm$1.28 &	0.039 & 6.75$\pm$1.33 \\
								
Hubble ACS  &  	GO-9796	 & F814W/I & 	2004 Feb 7 & 	2.99$\pm$0.19 & 0.47 & 	4.67$\pm$0.29 \\
Hubble ACS  &  	GO-9796	 & F555W/V & 	2004 Feb 7 & 	3.29$\pm$0.15 & 0.81 & 	7.06$\pm$0.31 \\
Hubble ACS  &  	GO-9796	 & F435W/B & 	2004 Feb 7 & 	3.87$\pm$0.10 & 1.08 & 	10.4$\pm$0.29 \\
Hubble ACS  &  	GO-9796	 & F330W/U & 	2004 Feb 7 & 	5.47$\pm$0.23 & 1.33 & 	18.6$\pm$0.74 \\
Hubble WFC3/IR  &  GO-12747  &  F160W/H & 2012 Sep 25 & 1.01$\pm$0.09 & 0.202 & 1.21$\pm$0.12 \\
								
XMM-Newton OM  & 0200980101 & 	UVW1 & 	2004 Sep 26 & 	3.79$\pm$0.32 & 1.51 & 	15.2$\pm$1.27 \\
XMM-Newton OM  & 0200980101 & 	UVM2 & 	2004 Sep 26 & 	$<$4.2 & 2.28 & $<$34.2 \\

XMM-Newton PN (obs. 1) & 0112521001 &  0.3 - 10 keV & 2002 Apr 10 & $\cdots$ & $\cdots$ & $\cdots$ \\
XMM-Newton PN (obs. 2) & 0693851701 &  0.3 - 10 keV & 2012 Nov 12 & $\cdots$ & $\cdots$ & $\cdots$ \\

\enddata

\tablecomments{(1):  Telescope and Instrument Name.  (2): Observation ID.  (3):  Filter for the photometric observations and the wavelength range for the spectral observations.  IRAC filters are microns. (4):  Date of observation. (5): Flux density of photometric observations in $\mu$Jy.  (6):  Total line of sight absorption for an extinction E(B-V)=0.26  (7): Extinction corrected photometric flux density in $\mu$Jy. }
\end{deluxetable}

\begin{deluxetable}{l|ccccccc}
\tablecolumns{7}  
\tablewidth{0pt} 
\tabletypesize{\tiny}
\setlength{\tabcolsep}{0.05in}
\tablenum{2} %\\               
\tablecaption{Spectral fits\label{table2}}   
\tablehead{
\colhead{Model} & \colhead{N$_H$} & \colhead{kT$_{in}$} & \colhead{$\Gamma$/$\tau_e$} & \colhead{R$_{in}$} & \colhead{kT$_{e}$} & \colhead{$\chi^2_{\nu}$} & \colhead{log L} \\
\colhead{} & \colhead{10$^{21}$ cm$^{-2}$} & \colhead{keV} & \colhead{} & \colhead{10$^{3}$ km} & \colhead{keV} & \colhead{} & \colhead{erg s$^{-1}$} \\
\colhead{(1)} & \colhead{(2)} & \colhead{(3)} & \colhead{(4)} & \colhead{(5)} & \colhead{(6)} & \colhead{(7)} & \colhead{(8)} \\
}
\startdata
\multicolumn{6}{l}{\tiny {X-ray + optical}}\\
\tableline

DISKIR obs 1     &  2.3$\pm$0.3  &	0.11$\pm$0.02        &  1.82$\pm$0.03          &  90.2$^{+13.7}_{-0.3}$ & 3.5$^{+0.9}_{-0.8}$ & 1.41/132 & 40.3 \\
DISKIR obs 2     &  2.45$^{+0.06}_{-0.22}$  &	0.13$\pm$0.01        &  1.61$\pm$0.02          &  80.3$^{+18.9}_{-0.2}$ & 1.59$^{+0.08}_{-0.07}$ & 1.45/144 &  40.6 \\

\tableline
\multicolumn{6}{l}{\tiny {X-ray only}}\\
\tableline

SIRF obs 1       &  2.35$\pm$0.07           & 2.1$^{+0.3}_{-0.2}$ & $\cdots$                & 0.49$\pm$0.07  & $\cdots$               & 1.27/131 &  40.3 \\
SIRF obs 2       &  2.8$^{+0.2}_{-0.1}$  & 1.25$^{+0.07}_{-0.06}$ & $\cdots$                & 0.66$\pm$0.03           & $\cdots$               & 1.19/143 &  40.6 \\
MCD+COMPTT obs 1 & 1.4$^{+0.8}_{-0.4}$   & 0.22$^{+0.07}_{-0.05}$ & 7$^{+1}_{-5}$  & $<$ 26.4                & 2.9$^{+0.9}_{-0.6}$ & 1.23/129 &  40.2 \\
MCD+COMPTT obs 2 & 1.6$^{+0.2}_{-0.3}$   & 0.7$\pm$0.2 & 12$^{+6}_{-5}$ & 4.6$^{+10}_{-3.5}$ & 1.7$\pm$0.2 & 1.21/141 &  40.4 \\

\enddata

\tablecomments{
Errors are limits indicate the 1$\sigma$ confidence regions.
(1): X-ray model, as described in Section~3.1.
(2): Intrinsic hydrogen column density. 
(3): Inner disk temperature.
(4): Photon index for DISKIR model and the optical depth of the Comptonizing component for the MCD+CompTT model.
(5): Inner disk radius for the accretion disk. For the DISKIR and MCD+COMPTT models these were calculated using the normalization.
For the SIRF model they were estimated using  the ``spherisation radius'' \citep[see][]{abo09} and the luminosity. See Section 3.1 for more details. 
(6): Electron temperature.
(7): Reduced $\chi^2$ values for the fit and the number of degrees of freedom. 
(8): Unabsorbed (intrinsic) luminosities between 0.3 and 10~keV.
}
\end{deluxetable}

\begin{figure}
\epsscale{0.9}
\plotone{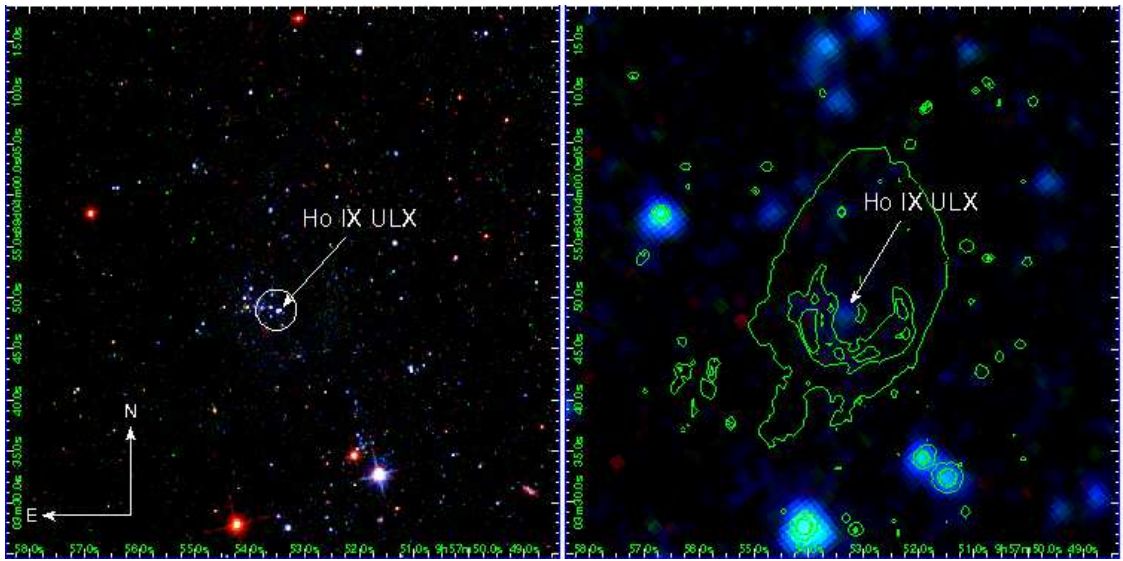}
\caption{
Left: HST ACS image in R (red), V (green) and B (blue) bands; right: IRAC image in 8.0 $\mu$m (red), 4.5 $\mu$m (green) and 3.6 $\mu$m (blue), with H$_{\alpha}$ contours from SUBARU FOCAS. The optical and IR counterpart is shown on both images, and the IRAC extraction aperture of 2 arcsec radius is shown on the HST image.}
\label{image}
\end{figure}

\begin{figure}
\epsscale{0.9}
\plotone{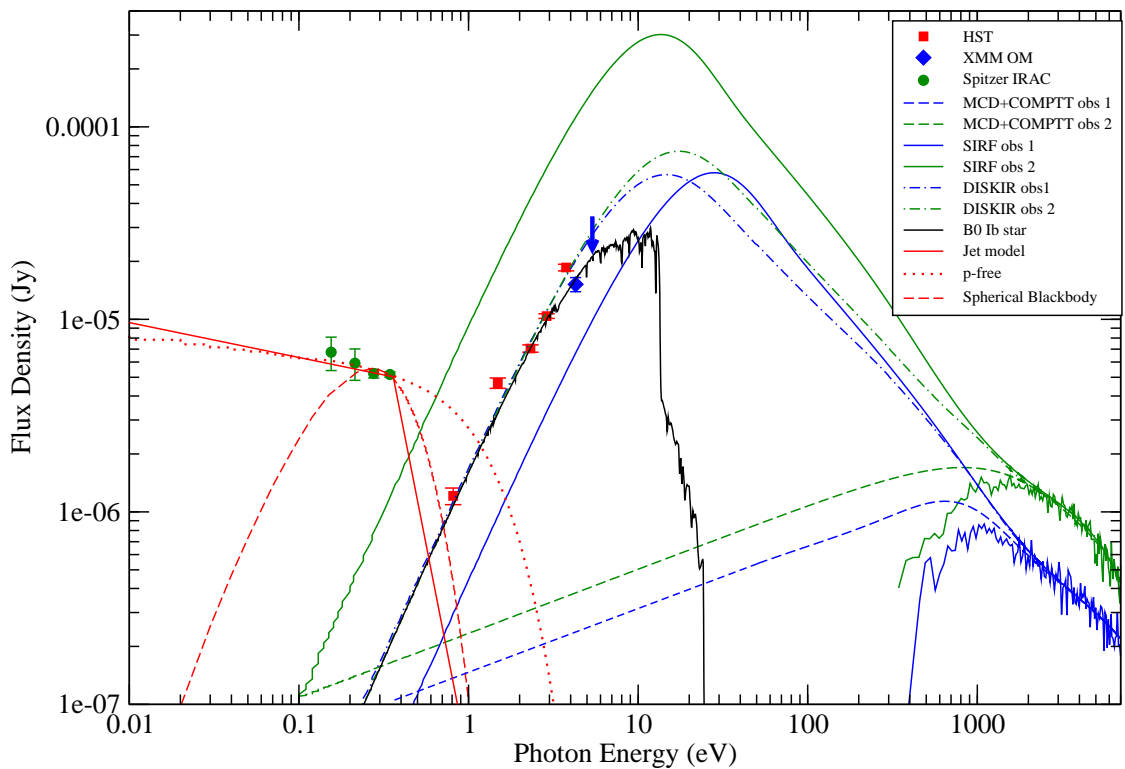}
\caption{
SED constructed in Section 3. The HST data is taken from \citet{gris11}, the XMM OM UV data from \citet{ber12}. The two representative X-ray datasets (XMM-Newton datasets 0112521101 and 0693851701 shown as obs1 in blue and obs 2 in green, respectively), are taken from \citet{lua16}.  Error bars are 1-sigma.} 
\label{sed}
\end{figure}


\begin{thebibliography}{}

\bibitem[Abolmasov et al.(2007)]{abo07} Abolmasov, P., Fabrika, S., Sholukhova, O., \& Afanasiev, V.\ 2007, Astrophysical Bulletin, 62, 36 
\bibitem[Abolmasov \& Moiseev(2008)]{abo08} Abolmasov, P., \& Moiseev, A.~V.\ 2008, \rmxaa, 44, 301 
\bibitem[Abolmasov et al.(2009)]{abo09} Abolmasov, P., Karpov, S., \& Kotani, T.\ 2009, \pasj, 61, 213 

\bibitem[Akeson et al.(2007)]{ake07} Akeson, R.~L., Rice, W.~K.~M., Boden, A.~F., et al.\ 2007, \apj, 670, 1240 
\bibitem[Bachetti et al.(2014)]{bach14} Bachetti, M., Harrison, F.~A., Walton, D.~J., et al.\ 2014, \nat, 514, 202 

\bibitem[Berghea et al.(2008)]{ber08} Berghea, C.~T., Weaver, K.~A., Colbert, E.~J.~M., \& Roberts, T.~P.\ 2008, \apj, 687, 471 
\bibitem[Berghea et al.(2010a)]{ber10a} Berghea, C.~T., Dudik, R.~P., Weaver, K.~A., \& Kallman, T.~R.\ 2010, \apj, 708, 354 (a)
\bibitem[Berghea et al.(2010b)]{ber10b} Berghea, C.~T., Dudik, R.~P., Weaver, K.~A., \& Kallman, T.~R.\ 2010, \apj, 708, 364 (b)
\bibitem[Berghea \& Dudik(2012)]{ber12} Berghea, C.~T., \& Dudik, R.~P.\ 2012, \apj, 751, 104 

\bibitem[Blandford K{\"o}nigl(1979)]{bla79} Blandford, R.~D., K{\"o}nigl, A.\ 1979, \apj, 232, 34 

\bibitem[Cardelli et al.(1989)]{card89} Cardelli, J.~A., Clayton, G.~C., \& Mathis, J.~S.\ 1989, \apj, 345, 245
\bibitem[Casares et al.(2014)]{cas14} Casares, J., Negueruela, I., Rib{\'o}, M., et al.\ 2014, \nat, 505, 378 
\bibitem[Colbert \& Mushotzky(1999)]{col99} Colbert, E.~J.~M., \& Mushotzky, R.~F.\ 1999, \apj, 519, 89 
\bibitem[Cseh et al.(2014)]{cseh14} Cseh, D., Kaaret, P., Corbel, S., et al.\ 2014, \mnras, 439, L1 

\bibitem[Done et al.(2004)]{done04} Done, C., Wardzi{\'n}ski, G., \& Gierli{\'n}ski, M.\ 2004, \mnras, 349, 393 
\bibitem[Fabbiano(1988)]{fabb88} Fabbiano, G.\ 1988, \apj, 325, 544 
\bibitem[Fabrika et al.(2015)]{fab15} Fabrika, S., Ueda, Y., Vinokurov, A., Sholukhova, O., \& Shidatsu, M.\ 2015, Nature Physics, 11, 551

\bibitem[Falcke \& Biermann(1995)]{fal95} Falcke, H., \& Biermann, P.~L.\ 1995, \aap, 293
\bibitem[Farrell et al.(2009)]{farr09} Farrell, S.~A., Webb, N.~A., Barret, D., Godet, O., \& Rodrigues, J.~M.\ 2009, \nat, 460, 73 
\bibitem[Fender \& Belloni(2004)]{fen04} Fender, R., \& Belloni, T.\ 2004, \araa, 42, 317 
\bibitem[Feng \& Soria (2011)]{fen11} Feng, H., \& Soria, R.\ 2011, \nar, 55, 166
\bibitem[Gerke et al.(2011)]{ger11} Gerke, J.~R., Kochanek, C.~S., Prieto, J.~L., Stanek, K.~Z., \& Macri, L.~M.\ 2011, \apj, 743, 176 

\bibitem[Gladstone et al.(2009)]{gla09} Gladstone, J.~C.,  Roberts, T.~P., \& Done, C.\ 2009, \mnras, 397, 1836
\bibitem[Greiner et al.(2001)]{gre01} Greiner, J., Cuby, J.~G., McCaughrean, M.~J., Castro-Tirado, A.~J., \& Mennickent, R.~E.\ 2001, \aap, 373, L37 
\bibitem[Gris{\'e} et al.(2011)]{gris11} Gris{\'e}, F., Kaaret, P., Pakull, M.~W., \& Motch, C.\ 2011, \apj, 734, 23 

\bibitem[Gris{\'e} et al.(2012)]{gris12} Gris{\'e}, F., Kaaret, P., Corbel, S., et al.\ 2012, \apj, 745, 123 

%\bibitem[Gris{\'e} et al.(2016)]{gris16} Gris{\'e} et al.\ 2016, private communication.
\bibitem[Hillen et al.(2015)]{hil15} Hillen, M., de Vries, B.~L., Menu, J., et al.\ 2015, \aap, 578, A40
\bibitem[Howell et al.(2006)]{how06} Howell, S.~B., Brinkworth, C., Hoard, D.~W., et al.\ 2006, \apjl, 646, L65 

\bibitem[King(2008)]{king08} King, A.~R.\ 2008, \mnras, 385, L113
\bibitem[La Parola et al.(2001)]{lap01} La Parola, V., Peres,  G., Fabbiano, G., Kim, D.~W., \& Bocchino, F.\ 2001, \apj, 556, 47
\bibitem[Makovoz \& Marleau(2005)]{mak05} Makovoz, D., \& Marleau, F.~R.\ 2005, \pasp, 117, 1113
\bibitem[Middleton et al.(2013)]{mid13} Middleton, M.~J., Miller-Jones, J.~C.~A., Markoff, S., et al.\ 2013, \nat, 493, 187

\bibitem[Middleton et al.(2015)]{mid15} Middleton, M.~J., Heil, L., Pintore, F., Walton, D.~J., \& Roberts, T.~P.\ 2015, \mnras, 447, 3243 
\bibitem[Miller \& Hodge(1994)]{mill94} Miller, B.~W., \& Hodge, P.\ 1994, \apj, 427, 656 

\bibitem[Miller(1995)]{mill95} Miller, B.~W.\ 1995, \apjl, 446, L75 
\bibitem[Motch et al.(2014)]{mot14} Motch, C., Pakull, M.~W., Soria, R., Gris{\'e}, F., \& Pietrzy{\'n}ski, G.\ 2014, \nat, 514, 198
\bibitem[Muno \& Mauerhan(2006)]{mun06} Muno, M.~P., \& Mauerhan, J.\ 2006, \apjl, 648, L135 
\bibitem[Liu et al.(2013)]{liu13} Liu, J.-F., Bregman, J.~N., Bai, Y., Justham, S., \& Crowther, P.\ 2013, \nat, 503, 500 
\bibitem[Luangtip et al.(2016)]{lua16} Luangtip, W., Roberts, T.~P., \& Done, C.\ 2016, \mnras
\bibitem[Pakull \& Mirioni(2002)]{pak02} Pakull, M.~W., \& Mirioni, L.\ 2002, arXiv:astro-ph/0202488
\bibitem[Pakull \& Mirioni(2003)]{pak03} Pakull, M.~W., \& Mirioni, L.\ 2003, Revista Mexicana de Astronomia y Astrofisica Conference Series, 15, 197

\bibitem[Pakull \& Gris{\'e}(2008)]{pak08} Pakull, M.~W., \& Gris{\'e}, F.\ 2008, A Population Explosion: The Nature \& Evolution of X-ray Binaries in Diverse Environments, 1010, 303 

\bibitem[Perez M.~\& Blundell(2010)]{per10} Perez M., S., \& Blundell, K.~M.\ 2010, \mnras, 408, 2 

\bibitem[Poutanen et al.(2007)]{pou07} Poutanen, J., Lipunova, G., Fabrika, S., Butkevich, A.~G., \& Abolmasov, P.\ 2007, \mnras, 377, 1187 


\bibitem[Rahoui et al.(2010)]{rah10} Rahoui, F., Chaty, S., Rodriguez, J., et al.\ 2010, \apj, 715, 1191 
\bibitem[Rahoui et al.(2011)]{rah11} Rahoui, F., Lee, J.~C., Heinz, S., et al.\ 2011, \apj, 736, 63 

\bibitem[Ramsey et al.(2006)]{ram06} Ramsey, C.~J., Williams, R.~M., Gruendl, R.~A., et al.\ 2006, \apj, 641, 241 
\bibitem[Revnivtsev et al.(2002)]{rev02} Revnivtsev, M., Sunyaev, R., Gilfanov, M., \& Churazov, E.\ 2002, \aap, 385, 904 

\bibitem[Soria \& Kuncic(2008)]{sor08} Soria, R., \& Kuncic, Z.\ 2008, American Institute of Physics Conference Series, 1053, 103
\bibitem[Stobbart et al. (2006)]{sto06} Stobbart, A.-M., Roberts, T.~P., \& Wilms, J.\ 2006, \mnras, 368, 397
\bibitem[Strohmayer \& Mushotzky(2003)]{str03} Strohmayer, T.~E., \& Mushotzky, R.~F.\ 2003, \apjl, 586, L61 
\bibitem[Sutton et al.(2013)]{sut13} Sutton, A.~D., Roberts, T.~P., \& Middleton, M.~J.\ 2013, \mnras, 435, 1758
\bibitem[Sutton et al.(2014)]{sut14} Sutton, A.~D., Done, C., \& Roberts, T.~P.\ 2014, \mnras, 444, 2415
\bibitem[Tao et al.(2012)]{tao12} Tao, L., Kaaret, P., Feng, 
H., \& Gris{\'e}, F.\ 2012, \apj, 750, 110 
\bibitem[Van Winckel et al.(1995)]{win95} Van Winckel, H., Waelkens, C., \& Waters, L.~B.~F.~M.\ 1995, \aap, 293,  

\bibitem[Vierdayanti et al.(2010)]{vier10} Vierdayanti, K., Done, C., Roberts, T.~P., \& Mineshige, S.\ 2010, \mnras, 403, 1206 
\bibitem[Walton et al.(2014)]{wal14} Walton, D.~J., Harrison, F.~A., Grefenstette, B.~W., et al.\ 2014, \apj, 793, 21
\bibitem[Webb et al.(2012)]{webb12} Webb, N., Cseh, D., Lenc, E., et al.\ 2012, Science, 337, 554 
\bibitem[Weng et al.(2014)]{weng14} Weng, S.-S., Zhang, S.-N., \& Zhao, H.-H.\ 2014, \apj, 780, 147 
\bibitem[Winter et al.(2006)]{win06} Winter, L.~M., Mushotzky, R.~F., \& Reynolds, C.~S.\ 2006, \apj, 649, 730 

\clearpage

%%%%%%%%%%%%%%%%%%%%%%%%%%%%%%%%%%%%%%%%%%%%%%%%%%%%%%%%%%%%%%%%%%%%%%%%%%%%%%%%%%%%%%%%%%%


\end{thebibliography}
\end{document}